\documentstyle[preprint,aps]{revtex}
\begin{document}
\draft

\title{QUARK CONFINEMENT AND DUAL REPRESENTATION 
IN 2+1 DIMENSIONAL PURE YANG-MILLS
THEORY}
\author{{\bf Sumit R. Das} and {\bf Spenta R. Wadia}}
\address{Tata Institute of Fundamental Research \\
Homi Bhabha Road, Bombay 400005, INDIA \\
e-mail: das@theory.tifr.res.in, wadia@theory.tifr.res.in}
\maketitle
\begin{abstract}
We study the quark confinement problem in 2+1 dimensional pure Yang-Mills
theory using euclidean instanton methods. The instantons are regularized
and dressed Wu-Yang monopoles. The dressing of a monopole is due to the
mean field of the rest of the monopoles.  We argue that such configurations
are stable to small perturbations unlike the case of singular, undressed
monopoles.  Using exact non-perturbative results for the 3-dim. Coulomb
gas, where Debye screening holds for arbitrarily low temperatures, we
show in a self-consistent way that a mass gap is dynamically generated in
the gauge theory. The mass gap also determines the size of the monopoles.
In a sense the pure Yang-Mills theory generates a
dynamical Higgs effect. We also identify the disorder operator of the
model in terms of the Sine-Gordon field of the Coulomb gas.
\end{abstract}
\narrowtext
\newpage

\section{Introduction}

The problem of quark confinement is one of the ``old'' unsolved
problems in theoretical physics. Despite intense activity over the
past two decades, and several approaches to the problem,
it is surprising how little we know about this phenomenon. Lattice
gauge theories, together with the theory of renormalization group
(which provides the basic conceptual framework for all other approaches
as well) is the only known {\em quantitative} and
reliable method of attacking this problem and Monte Carlo simulations
have indeed almost ``demonstrated'' confinement in pure gauge theories
in four dimensions. However there is the nagging feeling that
we can ``demonstrate'' confinement, even calculate relevant quantities
to some degree of accuracy - but we still dont ``understand'' confinement.
Despite impressive progress lattice methods remain a black box.
More specifically we do not understand in full generality
whether (if any) some specific type of gauge field configurations are
responsible for confinement and whether we can arrive at a consistent
picture of the vacuum of strongly coupled gauge theories.

Physical mechanisms for confinement are, of course, as old as the idea
itself. One of the most significant ideas, proposed by 
Mandelstam \cite{two}, 't Hooft \cite{three} and Nambu \cite{one}
is that of dual
superconductivity - that the gauge theory vacuum is a condensate of
magnetic monopoles. By the dual version of Meissner effect, quarks
would then be naturally confined. The formulation of this idea gave
rise to deep results concerning duality between electric and magnetic
fluxes and its implications to the phase diagram of general gauge
theories \cite{DISORDER}.  The idea is very convincing, 
and lattice results indeed seem
to support it : but is rather difficult to establish in usual
gauge theories.  Recently, however, there has been significant progress
in supersymmetric gauge theories where holomorphy and
global  properties of the
moduli space of vacua have been used to argue for
non-perturbative phenomena like quark confinement and chiral
symmetry breaking \cite{WITSEI}. This approach is
essentially Hamiltonian : one tries to obtain a picture of the vacuum wave
functional.

A complementary viewpoint stems from the Euclidean approach to the
problem. This is the idea that Euclidean instantons essentially
disorder the vacuum and lead to color confinement. The only known
successful implementation of this idea is the classic work of
Polyakov \cite{POLYAKOV}
who showed that confinement indeed occurs by this mechanism
in three dimensional $SU(2)$ Yang-Mills theory coupled to adjoint
representation Higgs field. The Higgs breaks the gauge group to
$U(1)$. The instantons of this model are the 't Hooft-Polyakov monopoles.
(In the three dimensional Euclidean theory these are of course not
solitons, but tunnelling configurations). When the mass scale of
this symmetry breaking - the mass of the $W$ boson, $m_W$ is large
compared to the mass scale set by the dimensional gauge coupling,
a dilute gas of monopoles provides a self-consistent picture. The
resulting monopole plasma leads to Debye screening.
%and at scales
%larger than the screening length it is described by a coulomb gas
%and hence a sine-gordon theory which has a mass gap.
Wilson loops
in the fundamental representation obey an area law and a careful
treatment shows that the adjoint representation Wilson loops obey
a perimeter law \cite{WADDAS}.
This is exactly what
one expects. The argument can be extended to the $SU(N)$ theory as
well.

In this paper we extend this approach to {\em pure} Yang-Mills
theory. This is a much more nontrivial system for several reasons.
Firstly, perturbation theory, though ultraviolet finite, is hopelessly
infrared divergent.
Secondly, because this theory has only one length
scale set by the gauge coupling $g$, one does not have the
luxury of having another length scale $m_W$ to enable a
controlled semiclassical approximation. The classical monopole
configurations are {\em singular} configurations in the continuum
limit and hence the renormalization of the monopole gas is very
nontrivial.

It is possible to regulate the singularity by modifying the
fields inside a ``core'' of some size $\lambda$. The classical
action now depends on $\lambda$. We do not vary the size $\lambda$
of the monopoles. Rather we treat it as a parameter of the theory
to be self consistently determined by the mass gap.

If the fluctuations around a regularised monopole solution are
decomposed in terms of representations of the direct product of spin
and color groups, the even parity S-wave fluctuations are unstable
\cite{YONEYA}. This is because the background magnetic field has a
long range Coulomb tail.
However the fluctuation problem in the Yang-Mills theory should
be performed not around a single monopole solution, but around
the neutral plasma of monopoles which populate the vacuum.
These monopoles have long range Coulomb interactions : other monopoles
affect the field around a typical monopole in a nontrivial way.
More significantly
the monopole positions and charges are fluctuating,
which make the {\em charge density} field of the monopoles a
dynamical variable. Such a fluctuation problem is difficult to
solve exactly.

In this work we address this question in an approximation guided
by the physics of the problem. We incorporate the fluctuations
of the charge density $\rho (x)$ by invoking known rigorous
results of the three dimensional Coulomb gas due to Brydges
\cite{BRYDGES}.  In \cite{BRYDGES} it has been shown that for a given
arbitrarily low temperature there is a chemical potential (fugacity) such
that the correlation of the density operator cluster.  In other words
there is Debye screening at all temperatures. The field around a given
charge thus decays exponentially over a {\em debye length} $l_D$ 
as opposed to a power law decay.

Our strategy is as follows. The aim is to show that Debye
screening is {\em self consistently} realized in the plasma of
magnetic monopoles.  The main complication is that the
fugacity of the Coulomb gas is itself a functional of the density
of monopoles $\rho(x)$.  However using the results of \cite{BRYDGES} we
can argue that for a given value $g^2$ of the gauge coupling there exists
a fugacity of the monopoles 
for which the plasma has a finite Debye length. 
The mean `magnetic field' is also screened with a fall off
given by the Debye length.  In this sense a monople configuration in a
plasma does not have a Coulomb tail and such a configuration which
incorporates this collective property we call a `dressed' monopole.  Small
fluctuations of the gauge field
around a dressed monopole are expected to be stable for
reasons similar to the Yang-Mills-Higgs theory.  
There the potential appearing in the stability operator decays
exponentially with a scale ${1\over m_W}$ due to contribution from
the Higgs field.
Hence in the pure Yang-Mills theory one seems to have a dynamical Higgs
effect that is produced by the monopole plasma. We also indicate how
the mass gap determines the size of the monopoles.

Finally we discuss the relation of our approach with the work
of 't Hooft \cite{DISORDER}. We give an explicit representation
of the disorder operators in $2 + 1$ dimensional Yang-Mills theory
and indicate that the dual theory is a $Z_2$ non-linear sigma
model.

Our discussion of the confinement problem gives a picture
of the dominant configurations in the Euclidean framework.
Feynman \cite{FEYNMAN} has given qualitative arguments for the ground
state wave function of this gauge theory in analogy with his work on the
roton spectrum in liquid helium. It would be
interesting to relate these two approaches.

The plan of this paper is as follows. In Section II we discuss
the topology of the vacuum in both the Georgi-Glashow model
and the pure gauge theory.
In Section III we briefly review the generation of mass gap and
confinement of $N$-ality in the
Yang-Mills Higgs system.
In Section IV we discuss self-consistent Debye screening of the monopole
plasma in the pure Yang-Mills theory. Section V is devoted to
the dual representation in terms of disorder operators.
Section VI is devoted to
conclusions.  In Appendix I we state the main result of
\cite{BRYDGES}.

\section{Topology of the gauge condition}

In this section we shall review the topological properties of the
vacuum of $SU(2)$ Yang-Mills theory in $2 + 1$ dimensions
following \cite{THOFT}. The physical degrees of freedom
are most transparent in a unitary type of gauge. For pure
gauge theories this is defined as follows. Consider some
local operator $X(x)$ which transforms according to the adjoint
representation of $SU(2)$. The unitary gauge is now defined
by
\begin{equation}
\label{one}
[X (x), \tau_3] = 0
\end{equation}
($\tau^i~,i = 1,2,3$ stands for the Pauli matrices).
This gauge condition retains the $U(1)$ generated by
$\tau^3$ as an unbroken symmetry. If the model contains
adjoint Higgs fields in addition to the gluons, $X$ may
be the Higgs field itself and (\ref{one}) is the 
conventional unitary gauge in such a Georgi-Glashow
model. In pure Yang Mills theory $X$ has to be constructed
out of the gauge fields alone.

We may write the matrix operator $X$ in the form
\begin{equation}
X = \lambda~I + \sum_{a=1}^3 \epsilon_a (x) \tau^a
\label{two}
\end{equation}
where $I$ is the identity matrix. Then the points $x_0$ where
\begin{equation}
\epsilon^a (x_0) = 0
\label{three}
\end{equation}
are singularities of the gauge condition (\ref{one}). As shown
in \cite{THOFT} these singularities are nothing but magnetic
monopoles with respect to the unbroken $U(1)$. In $3 +1$
dimensions these monopoles are ``particles'' which respresent
dynamical degrees of freedom (in this gauge) other than
the conventional fields. In $2 + 1$ dimensions these monopoles
are instanton configurations which will play a crucial role
in determining the properties of the vacuum.

In the Georgi-Glashow model the monopoles are given by the
well known 't Hooft-Polyakov solution, while for pure gauge
theory they are described by the Wu-Yang solution.

\section{Confinement in 2+1 dimensional
 Georgi Glashow model}

We now briefly discuss the salient features of the mechanism of quark
confinement in the Georgi-Glashow model (for
details see Ref. \cite{POLYAKOV} and \cite{WADDAS}). 

The model is described by the lagrangian density
\begin{equation}
\label{14}
{\cal L} = - {1 \over 2g^2}\; tr\; F_{\mu\nu}F^{\mu\nu}
 - tr (\nabla_\mu\Phi)^2
- {\lambda \over 4}\; \left(2 {\rm tr}~ \Phi^2 - v^2\right)^2
\end{equation}
where $A_\mu$ denotes the gauge field and $\Phi$ a Higgs field
in the adjoint representation of $SU(2)$.
The instantons of this model are 't Hooft-Polyakov monopoles.
For reasons to be discussed later we are interested in monopoles with
minimal charge :$q = \pm 1$  These are the 't Hooft-Polyakov
monopoles.  In the unitary gauge $[\Phi,\tau_3] = 0$
the fields are given by:
\begin{eqnarray}
\label{360}
{\tilde A}_\mu & = & {1 \over 2}\; \left[\matrix{q{\tilde A}^3_\mu 
& {\tilde A}^1_\mu - iq {\tilde A}^2_\mu \cr
{\tilde A}^1_\mu + iq{\tilde A}^2_\mu & -q 
{\tilde A}^3_\mu}\right] \nonumber \\
{\tilde \Phi} & = & q\; {H(vr) \over r}\; {\tau^3 \over 2}
\end{eqnarray}
where $q = \pm 1$ and
\begin{eqnarray}
\label{38}
{\tilde A}^1_\mu & = & = - {K (rm_W) \over r}\; 
\left[\hat\phi \cos\phi + \hat
\theta \sin \phi \right] \nonumber \\
{\tilde A}^2_\mu & = & {K(rm_W) \over r}\; 
\left[- \hat \phi \sin \phi + \hat
\theta \cos \phi \right] \\
{\tilde A}^3_\mu & = & - {1 \over r}\; \tan 
{\theta \over 2}\; [\hat \phi]_\mu
= D_\mu
\end{eqnarray}
where $(r, \theta, \phi)$ denote polar coordinates in space-times.
The functions $K(\xi),$ $H(\xi)$ obey well known differential
equations for (\ref{360}) - (\ref{38})
to be a classical Euclidean solution.  For $r
\gg m_W^{-1}$ one has $K(rm_W) \simeq 0$ and $H(vr) \simeq vr$.  $m_W^{-1}$
denotes the ``size'' of the monopole.

Note that given a configuration of monopoles and anti-monopoles,
the Weyl group changes {\em each} monopole into an anti-monopole
and vice-versa. In principle one may fix the gauge further to
remove this discrete degeneracy. We, however, prefer not to do so
and average over the Weyl group. Then one may freely perform a
sum over all the $q = \pm 1$.

As shown in \cite{POLYAKOV}, in the dilute gas approximation the
path integral may written as a grand canonical partition
function of a gas of monopoles:
\begin{equation}
\label{44}
Z = \sum^\infty_{N=0} {1 \over N!}\; J^NQ_N
\end{equation}
where
\begin{equation}
\label{45}
Q_N = \sum_{\{q_a\}} \int \prod dx_a \exp \left(- {2\pi \over g^2}
\sum {q_a q_b \over \vert x_a - x_b \vert}\right)
\end{equation}
and the fugacity $J$ is:
\begin{equation}
\label{46}
J = {16 \over \sqrt \pi}\; g^6s^{3/2}e^{-s} \left({\det D^2
\over \det -\partial^2}\right)^{-1/2} \left({\det \Delta_{FP} \over
\det - \partial^2}\right)
\end{equation}
$D^2$ denotes the small fluctuation operator around single monopole
field, $\Delta_{FP}$ is the Fadeev-Popov operator, and $s$ is the
action of a single monopole.

In (\ref{44}) - (\ref{46})
 we have treated the various zero modes of $D^2$ by the
standard procedure of collective coordinates \cite {GERSAK}.
The Weyl degeneracy of the unitary gauge condition has been accounted
for by summing over $q_a = 0, \pm 1$ for each space time point, as
noted earlier.

The Coulomb gas of eqns. (\ref{44}-\ref{46}) may be expressed as a massive
scalar field theory using the sine-Gordon transform:
\begin{equation}
\label{47}
Z = \int {\cal D}\chi (x) \exp \left[- {g^2 \over 32\pi^2} \int d^3x
\left\{(\nabla \chi)^2 - 2M^2 (1 - \cos \chi) \right\} \right]
\end{equation}
where
\begin{equation}
\label{48}
M^2 = {16\pi^2 J \over g^2}.
\end{equation}

Let us now examine self-consistency of the above scheme.
When $J$ is small, so that the scalar field theory is weakly coupled,
$M$ is precisely the mass gap.  Equation (\ref{47}) expresses the long
distance behavior of the theory in a dual representation.
The classical action of a single monopole has the form
\begin{equation}
s \sim {m_W \over g^2}
\end{equation}
When $m_W/g^2 \gg 1$, it follows from (\ref{46}) that $J/g^6 \ll
1$.  Further the ``classical'' piece in $J$ dominates over the
contribution from one-loop fluctuations.  Thus the scalar field theory
(\ref{47}) is indeed weakly coupled; $M$ is the mass gap and from 
(\ref{48})
$M/g^2$ is small.

Since $M$ is the mass gap, the correlation length $\xi = 1/M$.  Also,
$J$ denotes the probability of occurence of a single monopole and hence
the number of monopoles in a Debye volume of size $\xi$, $N_\xi$ is
given by
\begin{equation}
\label{50}
N_\xi = {J \over M^3}
\sim {g^2 \over M}  \gg 1
\end{equation}
where we have used (\ref{48}).

We immediately see that there is a large number of monopoles in one
Debye volume and hence the potential $\chi$ may be treated
classically. The situation here is
identical to the Debye-Huckel theory of electrolytes in the limit of high
temperatures where the Debye length is large and the smoothly varying
potential field $\chi$ satisfies the classical sine-Gordon equation.

The Wilson loop average is given by
\begin{equation}
\label{51}
\langle W(c) \rangle = {1 \over 2} \langle Tr\; P \exp \left(i\oint
A_\mu dx_\mu \right) \rangle
\end{equation}
where $A_\mu = A^a_\mu {\tau^a \over 2}$ when the external quarks are
in the fundamental representation of $SU(2)$ and $A_\mu = A^a_\mu T^a$
when the quarks are in the adjoint representation.  Here $T^a$ are the
generators of the adjoint representation.

The ``classical'' contribution from the
Wilson loop factors out:
\begin{equation}
\label{52}
\langle W(c) \rangle = \langle W(c)\rangle_{c\ell} [\omega(c)]_{qu}
\end{equation}
where $\langle W(c)\rangle_{c\ell} = {1 \over 2} \langle Tr\;P \exp
\left(i\oint \tilde A^a_\mu {T^a \over 2} dx_\mu\right)\rangle$, the
average being evaluated in the ensemble given by (\ref{47}).
$[\omega(c)]_{qu}$ obeys a perimeter law:
\begin{equation}
\label{53}
[\omega(c)]_{qu} \sim \exp (-\alpha P)
\end{equation}
where $\alpha$ is a constant and $P = {\rm perimeter~of~loop}$.

Let us first discuss the Wilson loop in the fundamental
representation. In the ensemble given by
(\ref{47}) one has
\begin{equation}
\label{61}
\langle W(c)\rangle^{\rm fund}_{c\ell} = \int {\cal D}\chi (y) \exp
[- {g^2 \over 32\pi^2M} \int d^3 y
(\nabla (\chi - {\eta \over 2})^2 -
2 (1 - \cos \chi))]
\end{equation}
Where $\eta(y)$ is the magnetic scalar potential due to a dipole layer of
unit strength on sheet $S$, which is the
solid angle subtended by the loop at the point $y$
In (\ref{61}) we have scaled the distances: $y \equiv Mx$.

Since $M$ is small, (\ref{61}) may be evaluated by stationary phase
approximation. The result is
\begin{equation}
\label{65}
\langle W(c)\rangle^{\rm fund}_{c\ell} = \exp (-\sigma A)
\end{equation}
where
\begin{equation}
\label{67}
\sigma = {g^2M \over 32\pi^2} \int^{+\infty}_{-\infty} dy_3
\left\{\left({d\bar\chi \over dy_3}\right)^2 + 2 \left(1 - \cos
\bar\chi\right)\right\} = {3g^2M \over 8\pi^2} \ldots
\end{equation}
and $A = {\rm area~of~the~loop~in~physical~units.}$. Then,
using (\ref{52}) it is easily seen that the Wilson loop
obeys an area law, indicating that quarks in the fundamental
representation are confined.

As shown in \cite{WADDAS} the Wilson loop in the adjoint
representation obeys a perimeter law instead. This happens
because there is no distinction between the quantum of
magnetic flux created by the Wilson loop and that created
by the monopoles in the vacuum.
The picture of confinement discussed in this section is valid for any
SU(N) gauge group.  The mechanism is essentially Debye screening in a
gas of ``non-Abelian'' monopoles belonging to the adjoint
representation of a dual group *SU(N) \cite{WADDAS}.

\section{Confinement in pure Yang-Mills theory}

In pure Yang Mills theory there is no Higgs field.
This has two consequences. First, in the absence of the second scale
(i.e. $m_W$), the perturbation expansion is infrared divergent
and the theory cannot
be defined perturbatively in the infinite volume limit.
Secondly the classical monopole solutions are Wu-Yang monopoles
\cite{WUYANG} which have zero size and infinite action. We will
regulate these monopoles by assigning a
size $\lambda$, which is explained in the next subsection. We will then
construct an expansion around a plasma of such monopoles and, as explained
in the introduction, argue that these monopoles are stable against
fluctuations.  

\subsection{The monopole solution}

We will consider monopoles which have some size $\lambda$.
The field due to single monopole at ${\vec x} = 0$ is given in the unitary
gauge by:
\begin{eqnarray}
\label{77}
\tilde A_\mu (x) & = & {1 \over 2} \; \left[ \matrix{q \tilde
A^3_\mu & \tilde
A^1_{\mu} -iq \tilde A^2_\mu \cr \tilde A^1_\mu + iq \tilde A^2_\mu & -q
\tilde A^3_\mu}\right]\nonumber \\
\tilde A^1_\mu (x) & = & - {K(r/\lambda) \over r} \; \left[\hat \phi
\cos\phi + \hat \theta \sin \phi\right]_\mu \nonumber \\
\tilde A^2_\mu (x) & = & {K(r/\lambda) \over r}\; \left[- \hat \phi
\sin\phi + \hat\theta\cos\phi\right]_\mu \\
\tilde A^3_\mu (x) & = & - {1 \over r}\; \tan\; {\theta \over 2}\;
[\hat\phi]_\mu = D_\mu \nonumber
\end{eqnarray}
in spherical coordinates.  Here $K(r/\lambda)$ is a structure function
regulating the fields at $r=0$.
The function $K(r/\lambda)$ goes to $1$ as $r \rightarrow 0$ as
follows
\begin{equation}
\label{89}
K(r/\lambda) \sim 1 - {r^2 \over \lambda^2} \;\; {\rm for}\; r
\rightarrow 0
\end{equation}
while at  $r = \lambda$, $K(r/\lambda) = 
K '(r/\lambda) = 0 $ and remains zero for
$r > \lambda$.  Furthermore $K'(r/\lambda)$ is continuous at
$r = \lambda$. As shown
by Banks, Myerson and Kogut \cite{BANKS}, one can choose such a
$K(r/\lambda)$ so that the configuration (\ref{77}) is a classical
solution.  The action of single monopole is
\begin{equation}
\label{80}
s = {4\pi \over g^2}
\int^\infty_0 dr \left[\left({dK\over dr}\right)^2 +
\left({K^2 -1 \over 2r^2}\right)\right]
\end{equation}

Note that the monopole field is abelian outside the monopole
core.
Consider a gas of such monopoles such that if the positions of
the monopoles are $x_a, x_b$ etc., one always has $\vert x_a
- x_b \vert > 2\lambda$.
The field
configuration in the regions outside the core of the monopoles is
approximated by
\begin{equation}
\label{81}
{\cal A}^{c\ell}_\mu (x) = \sum^N_{a=1} \tilde A_\mu (x-x_a)
\end{equation}
where $x_a$ denotes a monopole position and $N$ is the number of
monopoles.  In our self-consistent approach, we assume that
(\ref{81}) represents the dominant field configuration in the euclidean
path integral.
The total action of this gas is
\begin{equation}
\label{82}
S_{cl} = Ns + {2\pi \over g^2} \sum_{a\not= b} \; {q_a q_b \over |x_a -
x_b|}
\end{equation}

Finally we record the single monopole field configuration in the
radial gauge
\begin{equation}
A_\mu^a = -{\epsilon_{\mu a j}x^j \over r^2}(1-K(r/\lambda))
\label{radial}
\end{equation}

\subsection{The path integral}

The monopole configuration is used to evaluate the path integral by
the saddle point method.  One expands the field around ${\cal
A}^{c\ell}_\mu$ which is a classical solution outside the core.
The form of the solution inside the core is unimportant for our
purposes:
\begin{equation}
\label{83}
{\cal A}_\mu = {\cal A}^{c\ell}_\mu + g a_\mu
\end{equation}
The path integral may be formally written as:
\begin{equation}
\label{84}
Z = \int \prod_x d {\cal A}_\mu (x) \exp (- S_{c\ell}) \exp \left(-
\int aD^2a d^3x\right)
\end{equation}
where $S_{c\ell}$ is given by (\ref{82}).  Here $D^2$ is the stability
operator for small fluctuations:
\begin{equation}
\label{85}
\int a D^2a d^3 x \equiv \int {\cal L}'' ({\cal A}_{c\ell})\; a^2 d^3
x
\end{equation}
where ${\cal L}'' ({\cal A}_{c\ell})$ is the second functional
derivative of the Lagrangian density evaluated at ${\cal A}_{c\ell}$.

Equation (\ref{84}) as it stands is meaningless since $D^2$ has a zero
eigenvalue for each symmetry of the original Lagrangian.
The local gauge symmetry is fixed by requiring the fluctuations
to satisfy the background gauge condition
\begin{equation}
\label{87}
\nabla_\mu ({\cal A}^{c\ell}) a_\mu(x) = 0
\end{equation}
where $\nabla_\mu ({\cal A}^{c\ell})$ is the covariant derivative
evaluated at the configuration ${\cal A}^{c\ell}$:
\begin{equation}
\label{88}
\nabla_\mu ({\cal A}_{c\ell}) \equiv \partial_\mu + i \left[{\cal
A}_\mu^{c\ell},~~~ \right]
\end{equation}
This gives rise to the usual Fadeev-Popov determinant $({\rm det}~
\Delta_{FP})$. The zero modes arising from the breaking of the
global translation invariance and the $U(1)$ transformations
obeying the background gauge condition but nonvanishing at
infinity are replaced
by integration over corresponding collective coordinates. Finally
we have to sum over all $N$ with the standard division by $N !$.
One finally has the formal expressions
\begin{eqnarray}
\label{96}
Z & = & \sum^\infty_{N=0} {1 \over N!}\; Q_N \nonumber \\
Q_N & = & \sum_{\{q_a\}} \left({8 \over \sqrt \pi}\: s^{3/2}\right)^N
\int \prod^N_{a=1} dx_a e^{-S_{c\ell}}~{\cal J}
\end{eqnarray}
\begin{equation}
{\cal J} = \left({\det D^2 \over \det
-\partial^2}\right)^{-1/2} \left({\det \Delta_{FP} \over \det
-\partial^2}\right)
\label{966}
\end{equation}

\subsection{Instability of the Single Undressed Monopole}

In the usual semiclassical method the dilute instanton gas
\cite{CDG} is
noninteracting and one writes
\begin{equation}
\label{97}
\det D^2 = (\det d^2)^N
\end{equation}
where $d^2$ is the stability operator for the single monopole
configuration.

We will see, however, that the $d^2$ has negative eigenvalues
signifying the instability of a single monopole. In
the background gauge (\ref{87}) the operator $d^2$ has the form:
\begin{equation}
\label{98}
d^2 = \delta_{\mu\nu} \nabla_\alpha \left(\tilde A_{c\ell}\right) \:
\nabla^\alpha \left(\tilde A_{c\ell}\right) + i \left[\tilde F_{\mu\nu}^
{c\ell}, \;\;\right]
\end{equation}
where $\tilde A_{c\ell}$ is defined in (\ref{77}) and $\tilde
F_{\mu\nu}^{c\ell}$ is the corresponding field.  The Wu-Yang case
corresponds to $K=0$ in (\ref{77}). Following Yoneya \cite{YONEYA} we
shall use a spherical basis in the product space [(space-time) $\otimes$
(isospin space)].  In this product space $\tilde A_\mu^{c\ell}$ and
the fluctuations $a_\mu$ become tensors.  Then the unstable modes of
$d^2$ are given by
\begin{description}
\item[\rm (a)] Odd parity $S$ waves:
\begin{eqnarray*}
\phi_1 &=& a_{rr} \\
\phi_4 &=& {1 \over \sqrt 2} \left(a_{\phi\phi} +
a_{\theta\theta}\right)
\end{eqnarray*}
\item[\rm (b)] Even parity $S$ wave:
\[
\bar\phi_4 = {1 \over \sqrt 2}\; \left(\phi_{\phi\theta} -
\phi_{\theta\phi}\right)
\]
\end{description}
The tensor indices on $\phi$ refer to the abovementioned spherical
basis.  The corresponding eigenvalue equations are: (See Yoneya, Ref.
\cite{YONEYA})
\begin{equation}
\label{99}
\left(- {d^2 \over dr^2} + {3K^2 - 1 \over r^2}\right)\:
\left(r\bar\phi_4\right) = \alpha^2_e(r\bar\phi_4)
\end{equation}
\begin{equation}
\label{99a}
{2K^2 \over r^2}\:
\phi_1 - \sqrt 2 \left({K \over r}\: {d \over dr} +
{K - rK' \over r^2}\right)\: \phi_4 = \alpha^2_0 \phi_1
\end{equation}
\begin{equation}
\label{99b}
\sqrt 2 \left({1 \over r^2}\: {d \over dr}\: (rK) - {K - rK' \over
r}\right)\: \phi_1 + \left(- {d^2 \over dr^2} - {2 \over r}\: {d \over
dr} + {K^2 -1 \over r^2}\right)\: \phi_4 = \alpha^2_0 \phi_4
\end{equation}
\begin{equation}
\label{99c}
\left({d \over dr} + {2 \over r}\right)~\phi_1 = \sqrt 2\: {K \over r}\:
\phi_4
\end{equation}
For the Wu-Yang monopole $K=0$ and the instability is obvious from
(\ref{99}) and (\ref{99b}).  In this case both $r\phi_4$ and $r\bar\phi_4$
obey the equation
\begin{equation}
\label{100}
\left(- {d^2 \over dr^2} - {1 \over r^2}\right)\: \psi =
\alpha^2\psi,\:\: \psi = r\phi_4, r\bar\phi_4.
\end{equation}
The above equation is the Schrodinger equation for a particle in a
spherical potential $U_0(r) = -{\gamma \over r^2}$ with $\gamma =1$.
For such a potential it is known that when $\gamma > 1/4$ there are
bound states \cite{fifteen}.

When the structure function $K(r/\lambda) \not= 0$ the odd parity
$S$-waves become stable.  
Formally one can invert (\ref{99c}) to express $\phi_4$ in terms of
$\phi_1$ and insert this into (\ref{99a}) to obtain:
\begin{equation}
\label{101}
\left[ - \left({d^2 \over dr^2} + {2 \over r}\: {d \over dr}\right) +
{K^2+1 \over r^2} \right]  \left({r\phi_1 \over K}\right) = \alpha^2
\left({r\phi_1 \over K}\right)
\end{equation}
In the corresponding Schrodinger problem the ``potential'' is now
positive everywhere and hence $\alpha^2 > 0$.  Since one can now
rewrite (\ref{99c}) as
\begin{equation}
\label{102}
\phi_1 = {\sqrt 2 \over r^2} \int_0^r yK(y)\phi_4(y)dy
\end{equation}
stability of $\phi_4$ also follows.

However, regularising the field near $r=0$ does not remove the even
parity $S$-wave instability.  In fact (\ref{99}) has a potential that
becomes asymptotically $- {1 \over r^2}$ as $r \rightarrow \infty$.
For such a potential there are still an infinite number of negative
eigenvalues.  This result is independent of the detailed form of the
potential near $r=0$. Hence the single regularised monopole is unstable
and cannot be treated as a dominant configuration in the path integral.

It should be noted in the presence of Higgs fields
the potential is replaced by
\begin{equation}
\label{103}
{3K^2 + H^2 -1 \over r^2}
\end{equation}
where $H(vr)$ is the function in (\ref{360}).  As $r \rightarrow \infty$,
$H(vr) = vr + e^{-m_H r}$ (where $m_H = {{\sqrt{2\lambda}}v \over g}$ 
is the Higgs mass) , and hence the $1\over r^2$ Coulomb tail is
cancelled and we have a screened potential $e^{- m_H r} \over r^2$, which
removes the potential instability.

\subsection{Debye Screening in the Monopole Gas}

Our main point is that when the instantons are interacting,
the fluctuation problem cannot in general be factored
into $N$ copies of the fluctuation problem for an isolated
monopole. Rather one should consider the stability of the neutral
plasma of monopoles as a whole. This statement also applies to the
Yang-Mills-Higgs system in the previous section.
The main reason behind this is
that we have an integration over the positions of the monopole
and the charges, or equivalently a functional integration
over the charge density field. The effect of this averaging over
the charge density is very nontrivial. The results of \cite{BRYDGES}
show that the charge density field clusters so that the theory
of the density field generates a mass gap dynamically. We summarize these
results in Appendix I.

In fact the results of \cite{BRYDGES} mean that in the neutral plasma
the fluctuations of the magnetic field are bounded and the ${1\over
r^2}$ magnetic field of a single monopole is Debye screened to
${e^{-Mr} \over r^2}$, where $M(g^2,\lambda,z)$ is the
non-perturbative mass gap, which depends on the coupling $g^2$, the
monopole size $\lambda$ which is effectively the cut-off of the
Coulomb gas and the fugacity $z$.
Recall that the source of instability for a single isolated monopole is
the long range tail of the Coulomb potential. One might, therefore expect
that in the screened neutral plasma a ``dressed'' monopole whose
Coulomb tail has been screened can in fact be stable.

In the following we shall assume that the monopole gas is dilute.
Using translation invariance we focus on one monopole at $x = x_\alpha$
and its neighbourhood. We are thus considering the problem in
{\em the presence of a single source } at $x = x_\alpha$.
Recall that the fields outside the monopole
cores of size $\lambda$ are abelian. Thus, in the unitary gauge it
follows from
(\ref{81}) and (\ref{77}) that
outside the core of this monopole the field is
\begin{equation}
\label{104}
\tilde A^{\rm out}_{\mu} = \sum^N_{a=1} {1 \over 2}\: q_a
\pmatrix{D_\mu (x-x_a) & 0 \cr 0 & -D_\mu (x-x_a)}
\end{equation}
where we have set $x_N = x_\alpha$. Inside the core the effect
of the core field of the other monopoles can be ignored and we have
\begin{eqnarray}
\label{104b}
\tilde A^{\rm in}_{\mu}  & = & \sum^{N-1}_{a=1} {1 \over 2}\: q_a
\pmatrix{D_\mu (x-x_a) & 0 \cr 0 & -D_\mu(x-x_a)} \nonumber \\
& & + {1 \over 2}
\pmatrix{qD_\mu(x-x_\alpha) & {\tilde W}^{-}_\mu(x-x_\alpha) \cr
{\tilde W}^+_\mu(x-x_\alpha) & -qD_\mu(x-x_\alpha)}
\end{eqnarray}
where ${\tilde W}_\mu^\pm \equiv
{\tilde A}^1_\mu \pm iq {\tilde A}^2_\mu$ and ${\tilde A}^1_\mu,
{\tilde A}^2_\mu$ are as in (\ref{77}).  Introducing the charge density
\begin{equation}
\label{105}
\rho(x) \equiv \sum^N_{a=1} q_a \delta(x-x_a)
\end{equation}
and assuming that in the mean, for large $N$,
$\rho_N \simeq \rho_{N-1}$ we can rewrite
 (\ref{104b}) as
\begin{eqnarray}
\label{106}
A_\mu (x; x_\alpha,[\rho]) & = & \int d^3 y\: {\tau^3 \over 2}\:
\rho(y)D_\mu(x-y) \nonumber \\ + & &
\theta (\lambda-|x-x_\alpha |) \pmatrix{D_\mu(x-x_\alpha) & W^-_\mu (x -
x_\alpha) \cr W^+_\mu (x-x_\alpha) & - D_\mu(x-x_\alpha)}
\end{eqnarray}
In the above expression the sharp $\theta$ function may be replaced
by a smoother version. 

Debye screening means that {\em in the presence of a source} the
density $\rho(y)$ has a mean value $\bar \rho(y,x_\alpha)$, fluctuations
around which are small. $A_\mu (x; x_\alpha,[{\bar \rho}])$ then
represents a ``dressed'' monopole configuration. In our case this
``source'' is provided by the particular monopole at $x = x_\alpha$
in the plasma
and the statement pertains to the field in the neighbourhood of
this particular monopole.

The crucial point is that since Debye screening holds, we can 
 assume {\em self-consistently} that the gas of ``dressed'' monopoles is
weakly interacting, unlike the ``bare'' monopoles. The field around a
dressed monopole decay exponentially over a distance scale set by the
Debye screening length $l_D = {1\over M}$. If the average distance between
the monopoles is much larger than $l_D$ then the interaction
between such dressed monopoles vanishes and the operator $D^2$ has a
N-fold degeneracy. In other words the potential appearing in the
stability equation resembles $N$ far separated potential wells. In
this situation we have, using translation invariance,
\begin{equation}
{\rm det} D^2 \simeq ({\rm det} D^2 [\bar \rho])^N
\label{deter1}
\end{equation}
where $D^2[\bar\rho]$ now denotes the stability operator for
a {\em single} dressed monopole which is the same for any
monopole in the plasma. For finite distances between 
monopoles the exact degeneracy is lifted and eigenvalues of
$D^2$ organize themselves in bands. This would lead to
corrections to the result (\ref{deter1}) which may be
expanded in powers of ${l_D \over l_m}$ where $l_m$ denotes the
average distance between the monopoles. 

In this regard there is a difference between the pure Yang Mills
system and the Yang-Mills-Higgs system in the Higgs phase. As our
stability analysis in the previous subsection indicates (see e.g.
equation (\ref{103})) the presence of the Higgs field means that
the potential appearing in the stability operator around a single
monopole decays over length scales of the order of ${1\over m_W}$.
Since the debye length is much larger than ${1 \over m_W}$
corrections to the extensivity of the small fluctuation determinant
appear as powers of ${1 \over l_m m_W}$.

\subsection{The Sine-Gordon Transform and Dynamical generation
of Mass Gap}

We can now rewrite the theory in terms of a sine-Gordon model. The
path integral is written as
\begin{equation}
\label{112}
Z = \sum^\infty_{N=0} {1 \over N!}\: \left({8g^6 \over \sqrt \pi}\:
\bar s^{3/2}e^{-\bar s} \right)^N \sum_{\{q_a\}} \prod^N_{i=1} \int d\vec x_i
\exp \left(-{2\pi \over g^2} \sum_{a \ne b} {q_aq_b \over \vert x_a-x_b
\vert }\right)~ (\Theta [\bar \rho])^N
\end{equation}
where we have defined $\Theta[\bar\rho]$ (functional of the charge
densities) as
\begin{equation}
\label{113}
\Theta[\bar \rho] \equiv \left({\det D^2 [\bar\rho] \over \det -
\partial^2}\right)^{-1/2} \: \left({\det \Delta_{FP}(\bar\rho) \over \det -
\partial^2}\right)
\end{equation}
and ${\bar s}$ is the action for a single monopole.

It may be noted that we are dealing with a superrenormalizable theory.
Thus the expression (\ref{113}) is ultraviolet finite.

We then have
\begin{equation}
\label{901}
Z = \sum_{N=0}^{\infty} {1 \over N!}~\bar J^N \sum_{\{q_a\}}
\prod_{i=1}^N \int d \vec x_i~{\rm exp} \left( -
{2\pi \over g^2} \sum {q_a q_b \over \vert x_a - x_b \vert}
\right)
\end{equation}
where the mean fugacity $\bar J$ is given by the formula
\begin{equation}
\label{902}
\bar J = {8~g^6 \over {\sqrt{\pi}}}~\bar s^{{3 \over 2}}~
e^{-\bar s}~\Theta[{\bar \rho}]
\end{equation}

We can now perform the sine-Gordon transform as in section 3, and we
write (\ref{901}) as
\begin{equation}
\label{902a}
Z = \int {\cal D}\chi (x) \exp \left[- {g^2 \over 32\pi^2} \int d^3x
\left\{(\nabla \chi)^2 - 2\bar M^2 (1 - \cos \chi) \right\} \right]
\end{equation}
where
\begin{equation}
\label{902b}
\bar M^2 = {16\pi^2 \bar J \over g^2}.
\end{equation}

It is important to emphasize in accordance with the discussion of
\cite{BRYDGES}, that the quadratic term $\int d^3x (\nabla\chi)^2$ in
(\ref{902a}) must be understood in a regularised sense, so that the Coulomb
potential between the monopoles is valid only upto distances greater than
the core size $\lambda$.  In this sense $\lambda$ is the cut off (lattice
spacing) of the sine-Gordon theory.

\subsection{Stability of the dressed monopole}

We now discuss, in some more detail than previously, whether the function
$\Theta[{\bar \rho}]$ which is used in the definition of the fugacity $\bar J$
in (\ref{902}) is well defined. This issue is important because we
have already indicated in Section 4 that as we average over $\rho(x)$
the operator $D^2[\rho]$ has negative eigenvalues when $\rho$ corresponds
to a single isolated monopole.

We will now argue that $D^2[\bar\rho]$ is a positive operator. Recall
the form of $D^2[\bar\rho]$ in the unitary gauge, outside the
core of the dressed monopole which we can choose to be at
$x = 0$
\begin{equation}
\label{903}
D^2[\bar\rho] = - \delta_{\mu\nu} \nabla_\alpha (A(x,[\bar\rho]))
\nabla^\alpha (A(x,[\bar\rho])) + i \left[{\bar F}^{(3)}_{\mu\nu}(x,[\bar
\rho]),~~\right]
\end{equation}
where ${\bar F}^{(3)}_{\mu\nu}(x,[\bar\rho])$ is related to the
magnetic field by ${\bar B}_\lambda^{(3)} = \epsilon_{\mu\nu\lambda}
\bar F^{(3)}_{\mu\nu}$ and ${\bar B}_\lambda^{(3)}$ is given by
\begin{equation}
\label{904}
{\bar B}_\lambda^{(3)} = {\partial \over \partial x_\lambda}
\int d^3 y~{{\bar \rho}(y) \over | x - y|}
\end{equation}
Debye screening means that
\begin{equation}
\label{905}
{\bar B}_\lambda^{(3)} = {x^\lambda \over r^3}~e^{-Mr}~f(Mr)
\end{equation}
where the function $f(x)$ has the property that for $x >> 1$,
$f(x) \sim 1$,
and $M$ is the mass gap related to the Debye length $l_D$
by $M = {1 \over l_D}$

For a nonzero mass gap, the field outside the monopole core
cannot be transformed to the radial form of (\ref{radial}).
However close to the core and distances much smaller than
the Debye length $l_D$ the field is close to the single
monopole field outside the core and may be cast in the
radial gauge. Furthermore at distances much larger than the
Debye length, the field is close to zero and once again one
may cast the gauge potential in the radial gauge trivially
(i.e. with $K(r) =1$). As mentioned in Section 4 the
source of the even parity S-wave instability is the long
range Coulomb field of the monopole. Since screening cuts
off this Coulomb field and replaces it approximately by
an exponential, one expects stability.

The situation is in fact similar to that of the Yang-Mills-Higgs
system in some respects. Recall that the S-wave stability of the 't
Hooft-Polyakov monopole is ensured by the fact that the Higgs field
rises {\em exponentially} to $1$ beyond the core and cancels the
negative ${1 \over r^2}$ tail of the gauge field, preventing the
potential in the Schrodinger problem from being negative at large
distances.  In our problem the gauge field itself falls off
exponentially to zero and thus the potential in (\ref{99}) is positive
at large distances. In this sense we have a dynamical Higgs effect.

It is indeed possible to argue that the above argument for
stability is self-consistent. Recall that Debye screening means
that the form of the magnetic field away from the monopole core
is of the form \ref{905}. We are unable to determine the precise
form of the function $f(x)$, but we can parametrize it in the
following way. Inside the monopole core the magnetic field in
the unitary gauge follows from the form of the single monopole
vector potential
\begin{equation}
B_i^{(3)} = -{x^i \over r^3}(1-K(r\lambda)^2)
\label{newone}
\end{equation}
The function $K(x)$ is given by the solution in \cite{BANKS}.
Let us assume that this form of the magnetic field extends upto
$r = \alpha \lambda$ with some $\alpha > \lambda$. The magnetic
potential far outside the core is an exponential, as required
by Debye screening which gives
\begin{equation}
B_i^{(3)} = -{x^i \over r^3}(1+Mr)~e^{-Mr}
\label{newtwo}
\end{equation}
where $M$ is the mass gap generated in the plasma. We assume that
this form of the magnetic field extends from infinity upto the
point $r = \alpha \lambda$.  This gives the function $K$ 
in the region $r > \alpha \lambda$
\begin{equation}
K^2(r) = 1-(1+Mr)e^{-Mr}~~~~~~~~~~~~~~r > \alpha \lambda
\label{newthree}
\end{equation}
We further simplify the problem by replacing the function $K(r)$
inside the core by an exponential so that we get
\begin{equation}
K^2(r) = e^{-2r/\lambda}~~~~~~~~~~~~~~r < \alpha \lambda
\label{newfour}
\end{equation}
The potential which appears in the s-wave stability operator
is $V(r) = {3K^2 -1 \over r^2}$. This potential must be continuous
at $r = \alpha 1$ which determines $\alpha$ implicitly
through the equation
\begin{equation}
e^{-2\alpha} = 1 - (1 + M \alpha\lambda) e^{-M\alpha\lambda}
\end{equation}
It may be now seen that the potential $V(r)$ is negative in the
region ${1\over 2}{\rm log}~3~\lambda < r < r_2$ where $r_2$ may
be determined by the above considerations to be the solution of
$V(r) = 0$ for $r > \alpha \lambda$. This is approximately
${1.2 \over M}$

We now test whether such a potential can have bound states. This 
may be done by applying the Bargmann criterion for absence of
bound states
\begin{equation}
\int dr~ r V^{-} (r) < 1
\end{equation}
where $V^{-} (r)$ stands for the negative part of the potential.
A straightforward numerical integration then gives the result
that the Bargmann bound is satisfied for $\alpha < 1.38$. The
fact that the critical value of $\alpha$ came out to be greater
than one is an evidence for the self consistency of the picture.

The above considerations are rough : we have made several
simplifying assumptions. However these assumptions are dictated
by the physics of the problem. The numbers quoted above are
to be considered as indicative since they will change with 
different approximations to the function $f(r)$ and $K(r)$ inside
the core. However the above calculation gives a self consistent
argument in favor of the stability of monopoles in a neutral
plasma.

\subsection{Self consistent dynamical generation of monopole size}

As mentioned in Section the fugacity of the monopole gas depends
on ${\bar \rho}$ and hence it
has a dependence on the mass gap $M$ and the cutoff $\lambda$ of the
sine-Gordon theory. This means that the cutoff $\lambda$ is not
independent, but determined in terms of $M$ and $g^2$, i.e.
\begin{equation}
\label{876}
M = M(\lambda,g^2,z(\lambda,M))
\end{equation}
implies that
\begin{equation}
\label{877}
\lambda = {1 \over g^2} F({M \over g^2})
\end{equation}
where $F$ is a function obtained by inverting (\ref{876}). This
equation now fixes the monopole core size $\lambda$ as a function
of the mass gap $M$ in a self-consistent manner. The above considerations
must be considered as qualitative because the calculation of the
function $F$ in (\ref{877}) is beyond our present technology.

\section{Dual representation and the disorder operator}

We now relate the euclidean formalism in terms of disorder operators
introduced by 't Hooft \cite{DISORDER} . In the hamiltonian formalism
the Schrodinger picture disorder operator $\Phi_D ({\vec x_0})$ is
defined as an operator which creates a $Z_2$ magnetic vortex at the
point ${\vec x_0}$ in two dimensional space. 
More specifically it implements a singular gauge
transformation $\Omega^{[x_0]}$ which has the property that if we
consider a closed spatial loop $C$ parametrized by an angle $\theta$ one has
\begin{equation}
\Omega^{[x_0]}(\theta + 2\pi) = - \Omega^{[x_0]}(\theta)
\label{k1}
\end{equation}
when the loop $C$ encloses the point ${\vec x_0}$. If $C$ does
not enclose ${\vec x_0}$ the gauge transformation is single valued. 
Consider now the two point function of the Heisenberg picture
disorder operators
\begin{equation}
<\Phi_D^\dagger (x) \Phi_D (y)>
\end{equation}
Here $x$ and $y$ stand for three dimensional coordinates (including
the euclidean time). This two point function is then a sum over
all configfurations of the gauge fields which have a Dirac string
singularity along a line joining $x$ and $y$ with a monopole
of charge ${1\over 2}$ at the point $y$ and an antimonopole
of charge $-{1\over 2}$ at the point $x$. It is crucial that the
magnetic charges of these monopoles is half that of the monopoles
which populate the vacuum. They have magnetic charges so that
the Dirac string is {\em visible} by the lowest electrically
charged quarks which couple to the gauge field.

Repeating the steps which led to the sine-gordon representation
with the difference that we have two external magnetic sources
with charges $\pm {1\over 2}$ we easily get
\begin{equation}
<\Phi_D^\dagger (x) \Phi_D (y)> = <e^{{i\over 2}(\chi(y) - \chi(x))}>
\label{k3}
\end{equation}
the average on the right hand side in (\ref{k3}) 
being performed in the sine-gordon theory. 
A similar identification holds for all higher point correlation
functions of the disorder operators. Hence we can identify the
disorder operator with
\begin{equation}
\Phi_D (x) = e^{{i\over 2}\chi(x)}
\label{k4}
\end{equation}
In fact the sine-gordon action may be now written in terms
of $\Phi_D$ as
\begin{equation}
S = {g^2 \over 32 \pi^2}\int d^3 x [\partial_\mu \Phi_D^\dagger 
\partial_\mu \Phi + M^2((\Phi_D)^2 + (\Phi_D^\dagger)^2)]
\label{k5}
\end{equation}
upto an irrelevant constant. The field $\Phi_D$ is not
a conventional scalar field, since $\Phi_D^\dagger \Phi = 1$.
This non-linear $Z_2$ sigma model can be generalized to a
linear sigma model by the addition of the term
$\lambda\int d^3x(\Phi_D^\dagger \Phi -1)^2$ to the action 
(\ref{k5}). The action then exactly
has the form conjectured in \cite{DISORDER}.

The sine gordon theory thus is itself a dual representation
of the original Yang-Mills theory.
The action (\ref{k5}) has the global $Z_2$ symmetry $\Phi_D
\rightarrow \Phi_D^\dagger$ which is spontaneously broken leading
to magnetic disorder and confinement. This is simply the
symmetry $\chi \rightarrow - \chi$ of the sine-gordon
model. It is clear from the action (\ref{k5}) that the dimensionless
coupling constant is ${{\sqrt{M}} \over g}$. This is inversely
related to the gauge coupling $g$ as expected in a dual
formulation. The dual theory is weakly coupled when ${M \over g^2}$
is small. In this limit the minima of the potential (${\rm cos}~\chi$)
break the $Z_2$ symmetry spontaneously.

Finally we note that the above construction of the disorder
operators can be easily extended to $SU(N)$ gauge theories
following the treatment in \cite{WADDAS}.

\section{Conclusions}

We have argued that in 2+1 dimensional pure Yang-Mills theory Debye
screening in a gas of regularized and dressed Wu-Yang monopoles
provides a consistent picture of quark confinement.  We have used the
results of \cite{BRYDGES} that in a three dimensional Coulomb gas the
charge density field always clusters, leading to Debye screening even
for arbitrarily low temperatures. Our line of argument has been
self-consistent in nature, because Debye screening in turn implies a
screened magnetic field and hence the stability operator around a
single dressed monopole is expected to have no negative eigenvalues.
The mass gap thus obtained is non-perturbative and determines the
monopolesize self-consistently. A related issue is that the mean
configuration $\bar\rho (x)$ in the presence of a single monopole
source is in general non-classical and hence the associated scalar
potential $\bar\chi (x)$ does not satisfy a classical equation.  
Hence the explicit evaluation of the Wilson loops is not as
easily done.  However on general grounds the existence of a mass gap
leads to qualitative conclusions that are similar to the case of the
Yang-Mills-Higgs system. Finally we have obtained a representation
of the disorder operators of the theory in terms of the sine-gordon
field which leads to a dual representation of the gauge theory.

\newpage

\section{Appendix I}

In this appendix we state the main results of \cite{BRYDGES} on the 3-dim.
Coulomb gas.  Theorem 2.1 in Brydges \cite{BRYDGES}, adapted to our
notation states that: \\
Given any $c_1 > 0$, there exits $c_2 > 0$ such that for ${1 \over c_1} \leq
g^2\lambda$ and $z \leq {1\over 2} c^2_2g^6$ ($z$ is the fugacity) the
correlation functions of the density operator exits and clusters
exponentially, i.e. there exist strictly positive constants
$M(z,g^2,\lambda)$, $c'=c'(n')$ such that for $n_1 < n'$
\begin{equation}
\begin{array}{l}
\label{6.1}
|\langle \prod^{n_1}_{i=1} \rho(x_i) \prod^{n'}_{j=n_1+1} \rho(x_j + a)
\rangle - \langle \prod^{n_1}_{i=1} \rho(x_i)\rangle \langle
\prod^{n'}_{j=n_1+1} \rho(x_j)\rangle| \\
\leq c' \exp \left({- {\scriptstyle inf \atop 2 \leq n_1 < j \leq n'}|x_i -
x_j + a|}\cdot M\right)
\end{array}
\end{equation}
$M(z,g^2,\lambda)$ is the mass gap whose inverse is the Debye screening
length. In the limit $g^2 \lambda \rightarrow \infty$ one has the
classical Debye-Huckel limit ${M \over g^2} \rightarrow 0$.  If we apply
(\ref{6.1}) to the 2-point function of the density, for separation of the order
$\lambda$, the monopole core size, which is the lattice spacing for the
Coulomb gas we get
\begin{equation}
\label{6.2}
|\langle \Delta \rho(0) \Delta \rho (\lambda)\rangle|  \leq c' e^{-M\lambda}
\end{equation}
where $\Delta\rho (x) = \rho (x) - \langle \rho(x)\rangle$.
(\ref{6.2}) says that the fluctuations of $\rho(x)$ are bounded and finite.

\section{Acknowledgements}

We would like to thank J. Fr{\" o}hlich for comments and discussions
on a previous version of the manuscript. We also thank A. Dhar and
F. Hassan for discussions and A. Polyakov for a correspondence
clarifying some aspects of the previous version of the manuscript.

\end{document}